%% file: Facu_IDEA__1_.tex
\documentclass[pra,reprint,aps,sort&compress,amsmath,amssymb,superscriptaddress,floatfix]{revtex4-2}

\input{packages}
\begin{document}
	
\title{Entanglement formation in two-dimensional materials within microcavity}

\input{authors}	

\begin{abstract}
In this work, the entanglement generation between two hexagonal-lattice layers embedded in a microcavity is studied, accounting for both electromagnetic coupling and intrinsic spin-orbit interaction (SOI). Utilizing a short-time dynamical approach, we perform a perturbative Taylor expansion of the reduced density matrix to characterize the bipartite quantum correlations between the hexagonal layers. We demonstrate that the system undergoes a rapid transition from a localized product state in the conduction bands at $t = 0$ to a coherent superposition of valence and conduction band states. Our results indicate that the degree of entanglement is highly sensitive to the interlayer photon propagator, which contains the geometric ratios of the layer positions and the height cavity, and the specific Fermi energy and SOI signatures of the respective layers. We show the emergence of spacelike-separated quantum correlations in the ultra-short evolution regime, suggesting that heterostructures in cavities may be suitable to develop experiments for a deep understanding of spacelike-separated quantum effects. 
\end{abstract}

\maketitle

\section{Introduction}

It is well established that quantum systems with discrete spectra can harvest entanglement from the vacuum state of any relativistic quantum field\cite{reznik2}. 
Since the field ground state possesses intrinsic spatial entanglement, nonlocal correlations can emerge between space-like separated observables \cite{valentini, resnik, pozak}. When these quantum systems are modeled as local detectors interacting with the field, the correlations can be dynamically extracted through their mutual evolution, a process known as \emph{entanglement harvesting} \cite{resnik, valentini}. 
In the simplest scenario of two-level detectors, tracing over the field degrees of freedom leaves an entangled state for the detectors, even when they are causally disconnected. Although most analyses have focused on scalar and vector fields \cite{pozak}, this mechanism has found diverse applications in metrology \cite{salt}, cosmology \cite{stee, mar2, mar3}, quantum information processing \cite{jon, haw, al}, and energy teleportation \cite{hotta}. More recently, analogous processes have been proposed in the context of virtual particle dynamics in two-dimensional (2D) systems \cite{jsa1, jsaA, escudero0, fedeprb,kib3,kib2}, where vacuum fluctuations of the electromagnetic field mediate effective couplings between electronic states. In previous work, identical layered systems were considered \cite{facu2}. Here, we extend the framework to \emph{heterostructures} composed of distinct two-dimensional materials, allowing asymmetries in Fermi velocity, spin–orbit interaction (SOI), and cavity position. Lately, Unruh-de Witt detectors 
This generalization enables the study of how material-specific parameters modify the short-time generation of entanglement driven by vacuum fluctuations inside microcavities. 
This setting is experimentally relevant for heterostructures such as graphene/silicene or graphene/germanene interfaces \cite{hetero}, as well as for twisted bilayers where can be considered as systems with a Fermi velocity that depends on the relative rotation of the layers \cite{fedesolo} and for possible applications outside the physics \cite{stock}.

The remainder of this paper is organized as follows. In Sect. \ref{sec:Model} and \ref{III} we introduce the formalism to compute the time-dependent perturbation theory. In Sect. \ref{IV} the density operator is computed. In Sect. \ref{V} and Sect. \ref{VI} the entanglement entropy and concurrence entanglement measures are computed for the reduced density operator and present our results and discussions for different electronic states in both buckled monolayers. The concurrence and entanglement entropy are computed for different initial energies, angles, and SOI. Our conclusion follows in Sec. \ref{conclusions}.

\section{Model and Hamiltonian}\label{sec:Model}
We consider a regime in which interlayer electron tunneling and electron-electron interactions are negligible due to the spatial separation between the layers. This allows us to focus only on cavity-mediated interactions through virtual photon exchange. Each layer is described independently by an effective Dirac Hamiltonian, with a dependent Fermi velocity on the crystal and spin-orbit coupling. The coupling to the cavity field is treated within the minimal coupling prescription, assuming that the field is quantized inside a perfect planar cavity with discrete longitudinal modes.
To analyze the entanglement dynamics between two graphene layers, we have previously employed the density matrix formalism to compute entanglement measures (see Refs. \cite{pozak,mar1,mar2,mar3}). Following these established methods and in light of the formalism for the density matrix presented in Refs. \cite{jsa1,jsaA})], we consider a system composed of two parallel layers placed inside a planar microcavity. The electrons in each layer interact with the confined, quantized electromagnetic field. Each layer is allowed to have distinct parameters, such as Fermi velocity and SOC strength, reflecting possible asymmetry between the layers.
\begin{figure}[ht]
	\centering
	\includegraphics[scale=0.63]{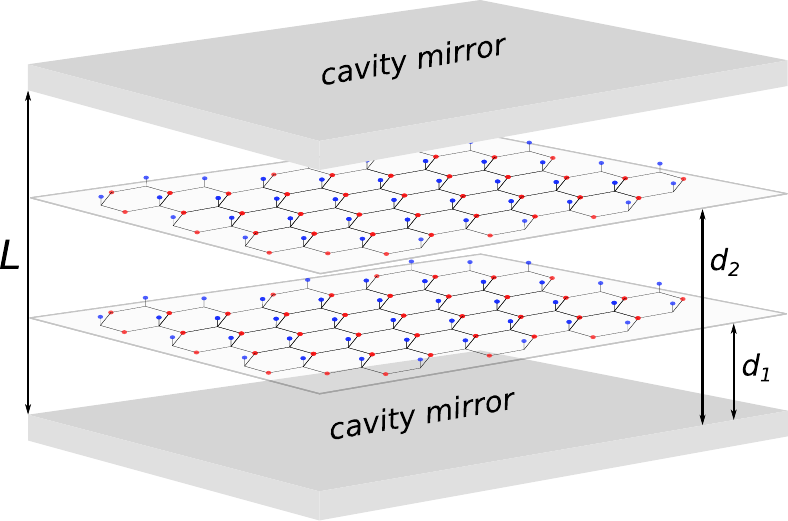}
	\caption{Schematic of the microcavity of length \( L \), with two hexagonal-lattice planes located at distances \( d_1 \) and \( d_2 \).}
	\label{fig:microcavity}
\end{figure}

The total Hamiltonian of the system reads
\begin{equation}
H = \sum_{i=1}^2 \left(H^{(i)}_0 + H^{(i)}_{\text{int}}\right) + H_F,
\end{equation}
where $H^{(i)}_0=v_f^{(i)}\boldsymbol{\sigma}^{(i)}  \cdot \vb{p}^{(i)} + \eta_i s_i \lambda_{so}^{(i)} \sigma_z^{(i)}  $ describes the free dynamics of electrons in the layer $i$, with different Fermi velocities $v_f^{(i)}$ in each layer \cite{peres-guinea,spen} and including spin-orbit interactions $\lambda_i$ \cite{par,zare1}, where $s_i=\pm1$ is the spin index and $\eta_i=\pm1$ is the valley index (for simplicity, enhancement of vacuum fluctuations are disregared and the Fermi velocity is not reduced \cite{escudero}). In turn,  $H^{(i)}_{\text{int}}=v_f^{(i)} \boldsymbol{\sigma}^{(i)}  \cdot \vb{A}^{(i)}  $ accounts for the minimal coupling between the electronic current and the electromagnetic field $\vb{A}$, and $H_F=\sum_{n,\textbf{q},\lambda} \hbar \omega_{n,\textbf{q},\lambda} \, a^{\dagger}_{n,\textbf{q},\lambda} a_{n,\textbf{q},\lambda} $ is the Hamiltonian of the free quantized field within the cavity \cite{kib, kakazu,jsa2018,jsa2}. The operators $a^{\dagger}_{n,\textbf{q},\lambda}$ and $ a_{n,\textbf{q},\lambda} $ satisfy the canonical bosonic commutation relations $\left[ a_{n,\textbf{q},\lambda}, a^{\dagger}_{n',\textbf{q}',\lambda'} \right] = \delta_{n,n'} \delta_{\textbf{q},\textbf{q}'} \delta_{\lambda,\lambda'}$
with photon frequencies $\omega_{n,\textbf{q}} = c \sqrt{|\textbf{q}|^2 + \left( \frac{n\pi}{L} \right)^2 } $ \cite{kakazu}, where $ \textbf{q}$ is the wavevector in the layer, $ c$  is the speed of light inside the cavity, and $ L $ is the length of the cavity.

\section{Interaction Picture and Perturbation Theory}\label{III}
We initialize the system in a factorizable state: the electrons in the different layers are both in the conduction band, and the cavity field is in the vacuum state. To study the correlations between the layers, the time-dependent perturbation theory can be applied considering the density operator formalism, where the interaction term $H^{(i)}_{\text{int}}$ is considered as a small perturbation compared to $H^{(i)}_0$ and \( H_F \), which describe the free evolution of the electrons and the cavity field, respectively. The initial state can be written as $\ket{\varphi(t_0)}_I = \ket{\nu_1} \otimes \ket{\nu_2} \otimes \ket{\Omega_0}$, where \( \ket{\Omega_0} = \ket{\Omega_0^+} \otimes \ket{\Omega_0^-} \) is the vacuum state of the electromagnetic field in circular polarizations \( \lambda = \pm \) and $\nu_i= \pm $ is the band index. To characterize the system's evolution, the initial density matrix can be computed as
\begin{equation}
	\begin{aligned}
		\rho_0 =& \ket{\varphi(t_0)}_I \bra{\varphi(t_0)} \\
		=& \ket{\Omega_0} \bra{\Omega_0} \otimes 
		\underbrace{ \ket{\nu_1} \bra{\nu_1} \otimes \ket{\nu_2} \bra{\nu_2} }_{ \rho_s = \ket{\nu_1, \nu_2} \bra{\nu_1, \nu_2} } 
		= \ket{\Omega_0} \bra{\Omega_0} \otimes \rho_s.
		\label{eq:rho0}
	\end{aligned}
\end{equation}

Using the time evolution operator $U_I(t,t_0)$ in the interaction picture, the time evolution of the density operator can be written as $\rho_T (t) = \ket{\varphi(t)}_I \bra{\varphi(t)} = U_I(t,t_0) \, \rho_0 \, U_I^\dagger(t,t_0)$. Using the Dyson expansion, up to the second order, the total density matrix reads (non-perturbative approach was given in \cite{brown} for more simpler quantum systems)
\[
\rho_T (t) = \rho_0 + \rho^{(1)} (t) + \rho^{(2)} (t)+ \cdots.
\]
%\begin{equation}
%	\rho_T = 
%	(1 + U^{(1)} + U^{(2)} + \cdots) \, \rho_0 \, 
%	(1 + U^{(1)\dagger} + U^{(2)\dagger} + \cdots),
%\end{equation}
%which can be grouped by order as:
%\begin{align}
%	\rho_T = \rho_0 \; 
%	 & + \; \underbrace{ U^{(1)} \rho_0 + \rho_0 U^{(1)\dagger} }_{\rho^{(1)}} \nonumber \\
%	&+ \underbrace{ U^{(1)} \rho_0 U^{(1)\dagger} + U^{(2)} \rho_0 + \rho_0 U^{(2)\dagger} }_{\rho^{(2)}} 
%	+ \cdots,
%	\label{eq:rhoT_expanded}
%\end{align}

%We work in the interaction picture with respect to the free Hamiltonian, and assume the system is initially in a product state between the field vacuum and a separable electronic state with support on two independent momenta. 
Since the excitations of the cavity field are not observed, then a partial trace of the total density operator with respect to the cavity field degrees of freedom must be computed. In turn, to probe interlayer correlations, the time-evolved reduced density matrix elements between electronic momentum eigenstates $\bra{\textbf{{k}}_1', \textbf{{k}}_2'} \rho_T(t) \ket{\textbf{{k}}_1, \textbf{{k}}_2}$ are considered, which describe fully delocalized electrons. Through a straightforward but lengthy calculation, the matrix elements $\rho_T$ between the electronic momentum eigenstates are
\begin{widetext}
\begin{equation}
	\begin{aligned}
		&\bra{\textbf{{k}}_1', \textbf{{k}}_2'} \rho_T(t) \ket{\textbf{{k}}_1, \textbf{{k}}_2} 
		= \delta_{\textbf{{k}}_1', \textbf{{k}}_1} \delta_{\textbf{{k}}_2', \textbf{{k}}_2} \rho_s - \frac{e^2 \gamma^2}{\hbar^2} 
		\sum_{i,j,\lambda} v_f^{(i)} v_f^{(j)} \Delta_{ij}(\vb{k}_i' - \vb{k}_i) 
		\delta_{\textbf{{k}}_i' - \textbf{{k}}_i, \textbf{{k}}_j - \textbf{{k}}_j'} \\
		& \int_{t_0}^{t} dt_1 \int_{t_0}^{t_1} dt_2 
		\left[
		\sigma_{-\lambda}^{(i)}(t_1) \sigma_{\lambda}^{(j)}(t_2) \rho_s - \sigma_{-\lambda}^{(i)}(t_1) \rho_s \sigma_{\lambda}^{(j)}(t_2) 
		e^{i\omega_n(t_1 - t_2)}
		+ \rho_s \sigma_{-\lambda}^{(i)}(t_1) \sigma_{\lambda}^{(j)}(t_2) 
		e^{i\omega_n(t_1 - t_2)}
		\right] 
	\end{aligned}
	\label{eq:final_rho_momentum}
\end{equation}
\end{widetext}

Here, $\sigma_{\lambda}^{(i)} (t)$ are the time-evolved Pauli matrices, $\gamma^2 =\hbar/(\epsilon_0 V)$, and the photon propagator is
\begin{equation}
	\Delta_{ij} = \sum_n 
	\frac{
		\sin\left( \frac{n\pi z_j}{L} \right) 
		\sin\left( \frac{n\pi z_i}{L} \right)
	}{
		\omega_{n, (\vb{k}_i' - \vb{k}_i)}
	}.
	\label{eq:Delta_def}
\end{equation}

The results obtained allow us to reformulate the model as an effective realization of the Unruh–DeWitt (UDW) detector model with a pair of two-level systems linearly coupled to a quantized field. Originally introduced to describe particle detection in quantum fields, the UDW model and its subsequent refinements \cite{strit} capture the essential mechanisms of entanglement harvesting. Recently, UDW detectors has found success in demonstrating quantum information channels with non-zero channel capacity between qubits and quantum fields. Considering graphene ribbons between detectors, the UDW model provide the necessary framework for experimentally realizable quantum computers with near-perfect channel capacity \cite{aspling}. Two-dimensional materials offer a natural platform for implementing such couplings, since their sublattice degree of freedom plays the role of an internal detector interacting locally with the field. Previous studies of double layer graphene in cavities \cite{jsa2018, facu1} have shown that photon-mediated correlations can lead to nonzero concurrence for times shorter than the light-crossing interval between layers, indicating the existence of spacelike-separated entanglement. 

\section{Density Matrix}\label{IV}
To study the formation of quantum entanglement between electrons, we consider the short-time regime and perform a Taylor expansion of the reduced density matrix elements $\bra{\vb{k}_1', \vb{k}_2'} \rho(t) \ket{\vb{k}_1, \vb{k}_2}$. Using the fundamental theorem of calculus and Leibniz's rule, it is not difficult to show that the first-order term vanishes identically. In turn, for the second-order term we obtain the following:
\begin{equation}
	\begin{aligned}
		\rho(t) &= \rho_s 
		- \frac{1}{2} t^2 
		\left( \frac{e\gamma}{\hbar} \right)^2 
		\sum_{i,j,\lambda} 
		v_f^{(i)} v_f^{(j)} \Delta_{ij} \\
		&\quad \times \left( 
		\sigma_{-\lambda}^{(i)} \sigma_{\lambda}^{(j)} \rho_s 
		- 2 \sigma_{-\lambda}^{(i)} \rho_s \sigma_{\lambda}^{(j)} 
		+ \rho_s \sigma_{-\lambda}^{(i)} \sigma_{\lambda}^{(j)} 
		\right)
	\end{aligned}
	\label{eq:taylor_second_order}
\end{equation}

% For convenience, we can separate the density matrix as
% \begin{equation}
% 	\rho = \rho_s + \rho_{\text{noise}} + \rho_{\text{signal}},
% 	\label{eq:rho_split}
% \end{equation}
% where $\rho_{\text{noise}} $ arises from terms with $i = j$ , that is, local noise contributions within each layer, and $ \rho_{\text{signal}} $ arises from $i \neq j $, encoding non-local correlations mediated by the photon propagator. The signal term is therefore responsible for encoding the interplane entanglement generated by virtual-photon exchange. Since the photon propagator has tails outside the light-cone, spacelike-separated entanglement generation is allowed between the layers. 

% \section{Change of basis}

To analyze the coupling between the conduction and valence bands mediated by circularly polarized photons, we can consider the action of $\sigma_\lambda$ on the basis of sublattice, where it can be verified that $\sigma_{\lambda} \ket{A,s} = \frac{1 - \lambda}{2} \ket{B,s}$ and $ \sigma_{\lambda} \ket{B,s} = \frac{1 - \lambda}{2} \ket{A,s}$.\\
To determine the action of \( \sigma_{\lambda} \) on the reduced density matrix \( \rho_s \), the sublattice states \( \{\ket{A,s}, \ket{B,s}\} \) can be expressed in terms of the eigenbasis \( \{\ket{+,s}, \ket{-,s}\} \) as
\begin{equation}
	\begin{aligned}
		\ket{A,s} &= \frac{1}{\Delta\chi} 
		\left( 
		\chi_+ \sqrt{1 + \chi_-^2} \ket{-} 
		- \chi_- \sqrt{1 + \chi_+^2} \ket{+} 
		\right), \\
		\ket{B,s} &= \frac{e^{-i\phi_{\vb{k}_i}}}{\Delta\chi} 
		\left( 
		- \sqrt{1 + \chi_-^2} \ket{-} 
		+ \sqrt{1 + \chi_+^2} \ket{+} 
		\right),
	\end{aligned}
	\label{eq:sublattice_to_band}
\end{equation}
with \( \Delta\chi = \chi_+ - \chi_- \), and 
\(
\chi_{\pm} = \frac{E_{\pm} - \Delta}{\epsilon} = \frac{\epsilon}{E_{\pm} - \Delta}
\). Since \( \sigma_+ = \ket{A,s}\bra{B,s} \) and \( \sigma_- = \ket{B,s}\bra{A,s} \), the following expressions can be obtained for \( \sigma_+ \) in terms of the band basis.
\begin{equation}
	\begin{aligned}
		\sigma_{\pm} = \frac{e^{\pm i\phi_k}}{\Delta\chi^2} 
		\Bigg\{ &
		\chi_- (1 + \chi_+^2) 
		\Big( \ket{-}\bra{-} - \ket{+}\bra{+} \Big) \\
		+ &\chi_{\mp} \sqrt{(1 + \chi_+^2)(1 + \chi_-^2)} \ket{+}\bra{-}  \\ 
            + &\chi_{\pm} \sqrt{(1 + \chi_+^2)(1 + \chi_-^2)} \ket{-}\bra{+}
		\Bigg\},
	\end{aligned}
	\label{eq:sigma_plus_band}
\end{equation}
Since terms like  $ \sigma_{-\lambda}^{(i)} \sigma_{\lambda}^{(j)} \rho_s $, 
$\sigma_{-\lambda}^{(i)} \rho_s \sigma_{\lambda}^{(j)}$, and $
 \rho_s \sigma_{-\lambda}^{(i)} \sigma_{\lambda}^{(j)} $ must be computed, it is necessary to determine the action of \( \sigma_\pm \) on \( \ket{\pm} \). Using the following identities,
\begin{equation}
	\begin{aligned}
		\text{(a)} &\quad \chi_- (1 + \chi_+^2) = -\chi_+ (1 + \chi_-^2), \\
		\text{(b)} &\quad \chi_+ \chi_- = -1, \\
		\text{(c)} &\quad \frac{\chi_+}{\chi_-} = -\chi_+^2, \\
		\text{(d)} &\quad \chi_- (1 + \chi_+^2) = -\Delta\chi, \\
		\text{(e)} &\quad \chi_-^2 = \frac{1}{\chi_+^2} = -\frac{\chi_-}{\chi_+} = 
		\frac{1 + \chi_-^2}{1 + \chi_+^2},
	\end{aligned}
	\label{eq:chi_identities}
\end{equation}
we finally obtain the following:
\begin{align}
    \sigma_{\lambda} \ket{\nu, s} &= 
	\frac{e^{i\lambda\phi_k}}{2\Delta\chi} 
	\sum_{\nu'} \nu' 
	\sqrt{ 
		\frac{1 + \chi_{k, \nu'}^2}{1 + \chi_{k, \nu}^2}
	}  \nonumber\\
	&\times \Big[
	(1 - \lambda) - (1 - \lambda) \chi_{k, -\nu'} \chi_{k, \nu}
	\Big]
	\ket{\nu', s}.
	\label{eq:sigma_on_band}
\end{align}
which gives the application of the Pauli matrices, written in the sublattice basis, on the eigenstates of the free Hamiltonian in each layer.

Combining the initial state and the perturbative contributions, the total density matrix can be expressed as

\begin{equation}
	\rho(t) = t^2
	\begin{pmatrix}
		\frac{1}{t^2} - \mathcal{L}_1 - \mathcal{L}_2 & \mathcal{B}_2 & \mathcal{B}_1 & \mathcal{M}^* \\
		\mathcal{B}_2 & \mathcal{L}_2 & \mathcal{N}^* & 0 \\
		\mathcal{B}_1 & \mathcal{N} & \mathcal{L}_1 & 0 \\
		\mathcal{M} & 0 & 0 & 0
	\end{pmatrix},
	\label{eq:rho_total}
\end{equation}

which is a similar result to the results previously presented in \cite{marti}, where the matrix coefficients read
\begin{align*}
	\mathcal{L}_i &= \left( \frac{e\gamma}{\hbar} \right)^2 \Delta_{ii} (v_f^{(i)})^2 
	\left( 1 - \frac{2}{\Delta\chi_i^2} \right), \\
	\mathcal{B}_i &= -\left( \frac{e\gamma}{\hbar} \right)^2 \Delta_{ii} (v_f^{(i)})^2 
	\left( \frac{\chi_{\nu_i} + \chi_{-\nu_i}}{\Delta\chi_i^2} \right), \\
	\mathcal{N} &= \left( \frac{e\gamma}{\hbar} \right)^2 
	\frac{v_f^{(1)} v_f^{(2)} \nu_1 \nu_2 \Delta_{12}}{\Delta\chi_1 \Delta\chi_2} N, \\
	\mathcal{M} &= \left( \frac{e\gamma}{\hbar} \right)^2 
	\frac{v_f^{(1)} v_f^{(2)} \nu_1 \nu_2 \Delta_{12}}{\Delta\chi_1 \Delta\chi_2} M.
\end{align*}

where 
\begin{align*}
	M &= -\chi_{\nu_1} \chi_{-\nu_2} e^{i(\phi_1 - \phi_2)} - \chi_{-\nu_1} \chi_{\nu_2} e^{-i(\phi_1 - \phi_2)}, \\
	N &= \chi_{-\nu_1} \chi_{-\nu_2} e^{-i(\phi_1 - \phi_2)} + \chi_{\nu_1} \chi_{\nu_2} e^{i(\phi_1 - \phi_2)}.
\end{align*}
The reduced electronic density matrix shows that the system, initialized as a product state, where both electrons are in the conduction band, develops coherence between the conduction and valence bands. The diagonal elements quantify the probability of occupation of different band configurations, showing that the matrix element $\ket{+, +} \bra{+, +}$ decreases in $t$ and simultaneously the other diagonal elements increase in $t$ obeying $Tr(\rho)=1$ of order $t^2$. Crucially, the off-diagonal elements capture quantum coherence among these configurations. These coherence are direct signatures of non-classical correlations and form the necessary basis for entanglement generation. In turn, the matrix element $\rho_{44}$ is zero, which implies that transitions to the valence band in each layer is not possible at order $t^2$. In the following section, we quantify this entanglement and determine the critical conditions for its emergence using entanglement entropy and concurrence. Should be stressed that the perturbative expansion is valid in the short-time regime, where the second-order Dyson term dominates the evolution. In particular, we focus on times smaller than the photon flight time between layers, where the presence of non-zero off-diagonal elements in the reduced density matrix signals spacelike-separated quantum correlations induced by vacuum fluctuations.

\section{Entanglement Entropy}\label{V}
Entanglement entropy quantifies the degree of quantum correlation between two parts of a composite system. We can compute it using the von Neumann entropy applied to the reduced density matrix of one of the subsystems. In this case, tracing out the degrees of freedom associated with the electron in the first layer, we obtain $\rho_2 = \mathrm{Tr}_1 \left( \rho \right)$. The validity of the perturbative expansion is grounded in the hierarchy of timescales. The density matrix elements $\rho_{ij}(t)$ are analytic functions of $t$ in the interaction picture, allowing a consistent Taylor expansion up to second order. The von Neumann entropy $S = -\text{Tr}(\rho \ln \rho)$ is then computed from the expanded density matrix. Although $S(\rho)$ itself exhibits nonanalytic behavior $\sim t^2 \ln t$ as $t \to 0$ (due to the $x \ln x$ term), this does not invalidate the expansion; it merely reflects the leading-order growth of entanglement, which can be rigorously extracted from the perturbative density matrix. Nevertheless the long-time limit cannot be taken, and the resulting entropies must be interpreted as instantaneous measures of correlation flow rather than equilibrium quantities. 

It is straightforward to show that the reduced density matrix \(\rho_2\) takes the form

\begin{equation}
	\rho_2 =
	\begin{pmatrix}
		1 - t^2 \mathcal{L}_2 & t^2 \mathcal{B}_2 \\
		t^2 \mathcal{B}_2 & t^2 \mathcal{L}_2
	\end{pmatrix},
	\label{eq:rho2_with_B}
\end{equation}

where the parameters are given by:
\begin{align*}
	\mathcal{L}_2 &= \left( \frac{e\gamma}{\hbar} \right)^2 \Delta_{22} (v_f^{(2)})^2 \left( 1 - \frac{2}{\Delta\chi_2^2} \right), \\
	\mathcal{B}_2 &= - \left( \frac{e\gamma}{\hbar} \right)^2 \Delta_{22} (v_f^{(2)})^2 \left( \frac{\chi_{\nu_2} + \chi_{-\nu_2}}{\Delta\chi_2^2} \right).
\end{align*}
Since the parameter \(\mathcal{B}_2\) decays rapidly for larger energies, it can be neglected in $\rho_{2}$, then the reduced density matrix becomes diagonal
\begin{equation}
	\rho_2 \approx
	\begin{pmatrix}
		1 - t^2 \mathcal{L}_2 & 0 \\
		0 & t^2 \mathcal{L}_2
	\end{pmatrix}.
	\label{eq:rho2_diagonal}
\end{equation}

To justify the diagonal approximation in last equation, we examine the relative magnitude of the coherence terms $\mathcal{B}_i$ compared to the population terms $\mathcal{L}_i$. Both coefficients scale identically as $t^2$ from the second-order perturbative expansion. However, their numerical prefactors differ: $\mathcal{L}_i$ contains the factor $1 - 2/\Delta\chi_i^2$, while $B_i$ contains $(\chi_{\nu_i} + \chi_{-\nu_i})/\Delta\chi_i^2$. For the parameter regime studied (electronic energies $\epsilon > 10\lambda$), numerical evaluation yields $|\mathcal{B}_i|/\mathcal{L}_i < 0.05$ for all materials considered. This justifies neglecting $\mathcal{B}_i$ in the leading-order analysis. The effect of including coherence terms is examined later in eqs. ( $\ref{eq:rho}$-$\ref{eq:entropy_with_B_expanded}$), where we show that their contribution to entropy appears at order $O(t^4)$ in the entropy expansion (not in the density matrix elements), due to the nonlinearity of the von Neumann functional $S$.

%In the primary analysis, we consider a diagonal approximation of the reduced density matrix by neglecting the coherence terms $\mathcal{B}_i$. This is justified by the rapid oscillatory nature of these terms at high energies, which leads to phase cancellation over the electronic bandwidth. Quantitatively, the contribution of $\mathcal{B}_i$ scales as $O(t^4 \ln t)$, whereas the entanglement generated by the population transfer $\mathcal{L}_i$ scales as $O(t^2 \ln t)$. Thus, in the limit $t \to 0$, the neglected coherences represent a higher-order correction that does not qualitatively alter the identified hierarchy of entanglement between the various hexagonal materials studied.

Applying the von Neumann entropy definition yields the entanglement entropy
\begin{equation}
\begin{aligned}
    S(\rho_2) &= - t^2 \mathcal{L}_2 \log (t^2 \mathcal{L}_2)
	- (1 - t^2 \mathcal{L}_2) \log (1 - t^2 \mathcal{L}_2).
	\label{eq:entropy_expression}
\end{aligned}
\end{equation}

\begin{figure}[ht]
	\centering
	\includegraphics[scale=0.7]{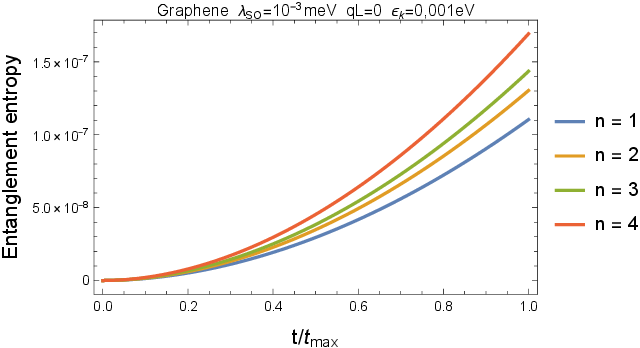}
	\includegraphics[scale=0.7]{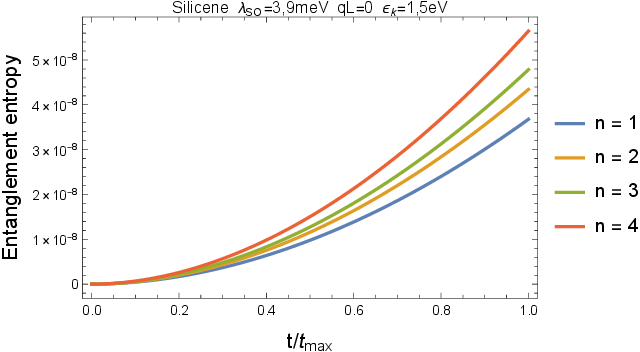}
	\caption{Entanglement entropy as a function of $t/t_{max}$ for graphene and silicene, respectively, considering different vibrational modes.}
	\label{fig:entropy_vs_time}
\end{figure}

The validity of the perturbative expansion requires $t^2 \mathcal{L}_i \ll 1$. For the parameters used in figure \ref{fig:entropy_vs_time} (graphene with $\lambda_{\text{SO}} = 10^{-3}$ meV, $\epsilon_k = 0.001$ eV, $n_{\text{max}} = 1$, cavity length $L = 1$ \(\mu\)m, and $d_2/L = 0.6$), we estimate $\mathcal{L}_2 \approx 2.3 \times 10^{17}$ s$^{-2}$. At the maximum time $t_{\text{max}} = 6.6 \times 10^{-10}$ s (light-crossing time), this yields $t_{\text{max}}^2 \mathcal{L}_2 \approx 0.10 \ll 1$, confirming that the perturbative regime is valid in the time range considered. The presence of spacelike-separated quantum entanglement between electrons in different layers can be studied using the entanglement entropy at subluminal time scales, i.e., before a photon can propagate from one layer to the other. Given that the photon propagator has tails outside the light cone, the contribution is nonzero even at very short times (\cite{propa}). For simplicity, considering that both layers are placed symmetrically with respect to the middle of the microcavity, the condition \(d_1/L = 1 - d_2/L\) is satisfied, where \(d_i\) is the position of layer \(i\) and \(L\) the cavity height. Setting \(d_2/L = 0.6\), we obtain \(d_1/L = 0.4\), so the time of flight of the photon between the layers is approximately \(t_{max} = 6.6 \times 10^{-10}\,\mathrm{s}\).\\
Figure \ref{fig:entropy_vs_time} shows that the entanglement entropy as a function of normalized time $t/t_{max}$ is non-zero for times smaller than the time of flight of photons between layers. Then entanglement is established for $t < L/c$. In both 2D materials, the entanglement entropy increases with the vibrational cut-off mode \(n_\text{max}\), due to the dependence of the photon propagator \(\Delta_{22}\) on both the vibrational modes and the position \(d_2/L\) of the second layer. In a realistic microcavity, this cut-off is naturally imposed by the plasma frequency of the metallic mirrors or the transparency regime of the dielectric Bragg reflectors, beyond which the electromagnetic field is no longer confined. For the numerical results presented, we chose $n_{max}$ such that the energy of the mode $\omega_{n_{max}}$ is large compared to the electronic energy scales ($\epsilon, \Delta$), ensuring that the virtual photon exchange is dominated by the resonant and near-resonant cavity modes.\\
\begin{figure}[ht]
	\centering
	\includegraphics[scale=0.7]{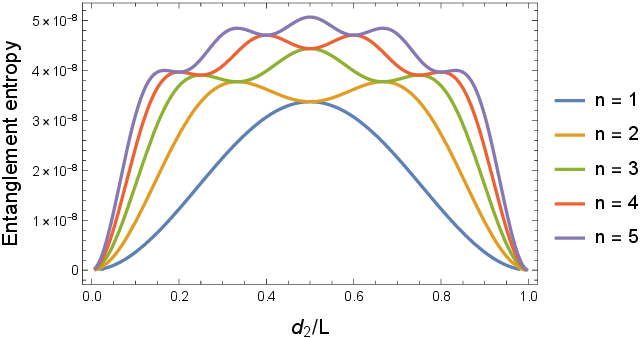}
	\includegraphics[scale=0.65]{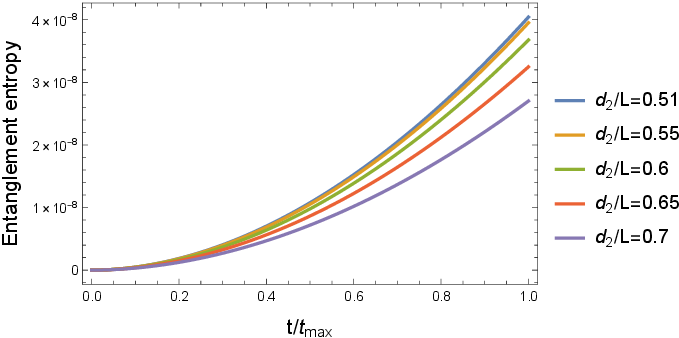}
	\caption{Top: Entanglement entropy as a function of \(d_2/L\) for various cutoff modes. Bottom: Entropy vs. $t/t_{max}$ for different interplanar distances in silicene.}
	\label{fig:entropy_vs_d2L}
\end{figure}
The upper plot in Fig. \ref{fig:entropy_vs_d2L} shows that the entanglement entropy increases for large vibrational cutoff, confirming that higher-mode contributions enhance non-local correlations. In turn, the functional form of the entanglement entropy, in terms of the normalized distance \(d_2/L\), shows maxima and minima associated with the discrete structure of the photon vibrational modes in the direction $z$, indicating a direct entanglement effect of the boundary conditions. Furthermore, the second figure reveals that the entanglement entropy increases as the interlayer distance decreases. This trend persists for any value of \(n_\text{max}\), although in this figure \(n_\text{max} = 1\) was used.\\
\begin{figure}[ht]
	\centering
	\includegraphics[scale=0.75]{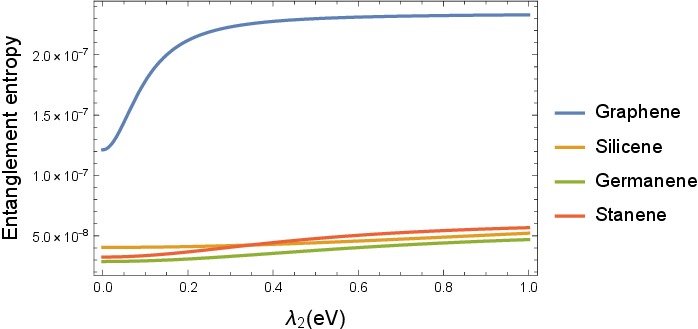}
	\caption{Entanglement entropy as a function of the SOI for different materials.}
	\label{fig:entropy_vs_lambda}
\end{figure}

Figure~\ref{fig:entropy_vs_lambda} explores the influence of spin-orbit coupling on the entanglement entropy, where it is shown that this quantity increases with SOI strength in all cases, where graphene shows the highest sensitivity. This indicates that a larger spin-orbit gap enhances the coupling between conduction and valence bands, thereby increasing the quantum correlations of electrons in different layers. Graphene, with the smallest intrinsic SOC ($\sim 10^{-3}$ meV), displays almost a discontinuous increment, while materials with larger SOC (stanene: $\sim 100$ meV) saturate at lower entropy values, reflecting a correlation between gap-enhanced coupling and reduced density of states near the Fermi level.
\\
\begin{figure}[ht]
	\centering
	\includegraphics[scale=0.75]{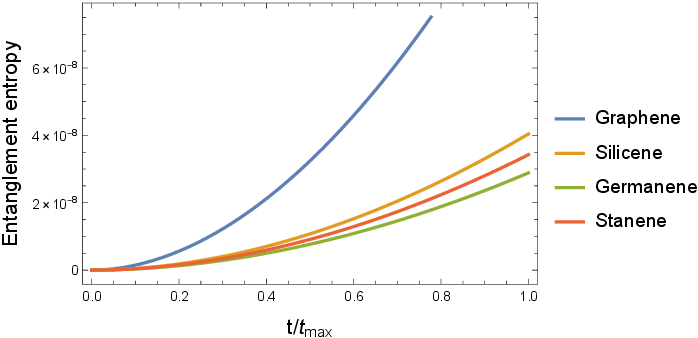}
	\caption{Entanglement entropy vs. $t/t_{max}$ for various materials: graphene, silicene, germanene, and stanene.}
	\label{fig:entropy_vs_material}
\end{figure}

Finally, Fig.~\ref{fig:entropy_vs_material} compares the entanglement dynamics between different 2D materials. In the spacelike-separated regime, graphene exhibits the largest entanglement, followed by silicene, germanene, and stanene. These results are obtained using a fixed vibrational cutoff mode $n_\text{max} = 1$ and identical interlayer separation. The reduced density matrices in Eqs. \eqref{eq:rho2_with_B}-\eqref{eq:rho2_diagonal} provide a complete description of how entanglement develops between the two layers. 
The von Neumann entropy of subsystem \(i\),
\begin{equation}
S_i(t) = - t^2 \mathcal{L}_i \log (t^2 \mathcal{L}_i)
- (1 - t^2 \mathcal{L}_i)\log (1 - t^2 \mathcal{L}_i),
\label{eq:entropy_rho2_full}
\end{equation}
encodes the degree of mixing induced by the interlayer coupling. 
In the short-time limit, it follows that
\begin{equation}
S_i(t) \simeq t^2 \mathcal{L}_i \left[ 1 - \log (t^2 \mathcal{L}_i) \right] + \mathcal{O}(t^4),
\label{eq:entropy_smallt_expansion}
\end{equation}
indicating that entanglement generation is a second-order process in the effective interaction strength \(\mathcal{L}_i\).
The instantaneous rate of entropy growth,
\begin{equation}
\Gamma_i = \frac{dS_i}{dt} = 2t \mathcal{L}_i \left[ 1 - \log(t^2 \mathcal{L}_i) \right],
\label{eq:entropy_rate}
\end{equation}
defines the flux of quantum information exchanged between the subsystems, revealing that entanglement increases coherently with a time-dependent rate proportional to both \(t\) and \(\mathcal{L}_i\). When the two layers are non-identical (\(v_f^{(1)} \neq v_f^{(2)}\) or \(\Delta_{11} \neq \Delta_{22}\)), the asymmetry in the local parameters leads to unequal entanglement generation rates. The difference in entropies, $\Delta S = S_1 - S_2$, quantifies this imbalance and determines the direction of quantum information flow. To first order in the asymmetry,
\begin{equation}
\Delta S \simeq 2t^2 \left[ (v_f^{(1)})^2 \Delta_{11} - (v_f^{(2)})^2 \Delta_{22} \right] 
\left( \frac{e\gamma}{\hbar} \right)^2,
\label{eq:entropy_asymmetry}
\end{equation}

The mutual information relation $I_{12} = S_1 + S_2 - S_{12}$ simplifies to $I_{12} = S_1 + S_2$ only if the full electronic state (after tracing the cavity field) remains pure, i.e., $S_{12} = 0$. To verify this, we compute the purity $\mathcal{P} = \text{Tr}(\rho^2)$ of the density matrix Eq. \ref{eq:rho_total}. Direct calculation yields:
\begin{equation}
\mathcal{P} = \text{Tr}(\rho^2) = 1 - 2t^2(\mathcal{L}_1 + \mathcal{L}_2) + O(t^4).
\end{equation}
For a pure state, $\mathcal{P} = 1$. The deviation from unity is $O(t^2)$, which reflects the second-order mixing introduced by the cavity-mediated interaction. However, the von Neumann entropy of the full electronic system vanishes to leading order in $t^2$: $S_{12} = -\text{Tr}(\rho \ln \rho) \approx 2t^2(\mathcal{L}_1 + \mathcal{L}_2)[1 - \ln(t^2 \mathcal{L}_i)] + O(t^4 \ln t)$. Computing this explicitly using the eigenvalues of $\rho$ (which are $1 - 2t^2(\mathcal{L}_1 + \mathcal{L}_2) + O(t^4)$) confirms $S_{12} \approx 0$ to leading order, since the logarithmic divergence cancels against the small eigenvalues. Therefore, within the perturbative regime, $I_{12} \approx S_1 + S_2$ is justified as a measure of total correlation.
In the symmetric case (\(\mathcal{L}_1=\mathcal{L}_2\)), one has \(S_1=S_2=S(t)\) and therefore $I_{12} = 2S(t)$, showing that each subsystem carries exactly half of the total information encoded in the bipartite entangled state. Deviations from this symmetry (\(\Delta S \neq 0\)) indicate an entropy flow between the two layers, corresponding to a directed exchange of quantum information while preserving the unitarity.

Regarding the effects of the coherence term \(\mathcal{B}_i\) in the reduced density matrix, the partial trace reads
\begin{equation}
\rho_i =
\begin{pmatrix}
1 - t^2\mathcal{L}_i & t^2\mathcal{B}_i\\[3pt]
t^2\mathcal{B}_i & t^2\mathcal{L}_i
\end{pmatrix}.
\label{eq:rho}
\end{equation}

The eigenvalues of \(\rho_i\) are then
\begin{equation}
p_{\pm}=\frac{1}{2}\!\left[1\pm\sqrt{(1-2t^2\mathcal{L}_i)^2+4t^4\mathcal{B}_i^2}\,\right],
\label{eq:exact_eigs}
\end{equation}
and their short-time expansion reads
\begin{equation}
p_- = t^2\mathcal{L}_i - t^4\mathcal{B}_i^2 + \mathcal{O}(t^6),\qquad
p_+ = 1 - t^2\mathcal{L}_i + t^4\mathcal{B}_i^2 + \mathcal{O}(t^6).
\label{eq:eig_series}
\end{equation}
The entropy follows as
\begin{equation}
S_i(t) = t^2\mathcal{L}_i\!\left[1-\log(t^2\mathcal{L}_i)\right]
+ t^4\!\left[\frac{\mathcal{L}_i^2}{2}+\mathcal{B}_i^2\log(t^2\mathcal{L}_i)\right]
+\mathcal{O}(t^6).
\label{eq:entropy_with_B_expanded}
\end{equation}
The second term shows that coherence contributions \(\propto \mathcal{B}_i^2\) reduce the entropy for small \(t\), meaning that local coherences tend to counteract the mixing produced by the population terms. Their effect on mutual information appears only at \(\mathcal{O}(t^4)\), leading to small negative corrections proportional to \(\mathcal{B}_1^2\log(t^2\mathcal{L}_1)+\mathcal{B}_2^2\log(t^2\mathcal{L}_2)\).

Within the validity of the perturbative expansion, the hierarchy of effects implies that coherence corrections  $\propto \mathcal{B}_i^2\,\mathcal{O}(t^4)$. Beyond this regime, higher-order terms or resummation schemes (e.g., master-equation approaches) are required to describe the long-time behavior and possible oscillatory or dissipative dynamics. 

\section{Concurrence}\label{VI}
Concurrence is one of the most widely used measures to quantify the entanglement between two quantum systems \cite{horo}. It is particularly suited for systems composed of two qubits, such as electrons with spin $1/2$. For a two-qubit system described by a density matrix \(\rho\), the concurrence \(C(\rho)\) is defined as (\cite{woo} and \cite{woo2})
\begin{equation}
	C(\rho) = \max\left(0, \lambda_1 - \lambda_2 - \lambda_3 - \lambda_4\right),
	\label{eq:concurrence_def}
\end{equation}
where \(\lambda_1, \lambda_2, \lambda_3, \lambda_4\) are the square roots of the eigenvalues, in decreasing order, of the matrix \(\rho (\sigma_x \otimes \sigma_x) \rho^* (\sigma_x \otimes \sigma_x)\). Here, \(\sigma_x\) is the Pauli matrix along the \(\hat{x}\) direction, and \(\rho^*\) denotes the complex conjugate of the density matrix. Using the total density matrix, we compute the matrix \(\rho (\sigma_x \otimes \sigma_x) \rho^* (\sigma_x \otimes \sigma_x)\) and its eigenvalues, which are \(|\mathcal{M}|^2\) and \(|\sqrt{\mathcal{L}_1 \mathcal{L}_2} \pm |\mathcal{N}||^2\). The derivation of these eigenvalues for the density matrix structure Eq.~(10) is provided in Appendix~A. Briefly, the matrix $\rho(\sigma_x \otimes \sigma_x)\rho^*(\sigma_x \otimes \sigma_x)$ has a block-diagonal structure after applying the spin-flip operation, which allows explicit diagonalization. The resulting eigenvalues depend only on the off-diagonal elements $M$ and $N$, as expected for X-state density matrices~\cite{rau}.
 From Eq.~\eqref{eq:concurrence_def}, the concurrence simplifies to
\begin{equation}
	C = 2\left(|\mathcal{N}| - |\mathcal{M}|\right),
	\label{eq:concurrence_simplified}
\end{equation}
This expression depends only on elements of the density matrix $i\neq j$, which means that it is influenced by the information flow between layers and thus directly determined by the interlayer photon propagator \(\Delta_{12}\). This dependence on off-diagonal matrix elements reflects the coherence between different band configurations, analogous to the coherence-based entanglement observed in other cavity QED systems (\cite{jsa1}, \cite{jsa2} and \cite{li}). Replacing the expressions for \(\mathcal{N}\) and \(\mathcal{M}\), we obtain
\begin{widetext}
\begin{equation}
	\begin{aligned}
		C =\ & \underbrace{\frac{(e\gamma)^2 v_f^{(1)} v_f^{(2)}}{\hbar^2}}_{\zeta} t^2 \frac{\epsilon_1 \epsilon_2 \Delta_{12}}{\sqrt{2} \sqrt{\epsilon_1^2 + \lambda_1^2} \sqrt{\epsilon_2^2 + \lambda_2^2}} \\
		& \times \left(
		-\sqrt{1 + \frac{\lambda_2^2}{\epsilon_2^2} + \frac{2\lambda_1 \left( \epsilon_2^2 \lambda_1 + 2 \lambda_2 (\lambda_1 \lambda_2 - \sqrt{\epsilon_1^2 + \lambda_1^2} \sqrt{\epsilon_2^2 + \lambda_2^2}) \right)}{\epsilon_1^2 \epsilon_2^2} + \cos(2\Delta\phi)}\right. \\
		& \left. + \sqrt{1 + \frac{\lambda_2^2}{\epsilon_2^2} + \frac{2\lambda_1 \left( \epsilon_2^2 \lambda_1 + 2 \lambda_2 (\lambda_1 \lambda_2 + \sqrt{\epsilon_1^2 + \lambda_1^2} \sqrt{\epsilon_2^2 + \lambda_2^2}) \right)}{\epsilon_1^2 \epsilon_2^2} + \cos(2\Delta\phi)}
		\right)
	\end{aligned}
	\label{eq:concurrence_full}
\end{equation}
\end{widetext}

\begin{figure}[ht]
	\centering
	\includegraphics[scale=0.75]{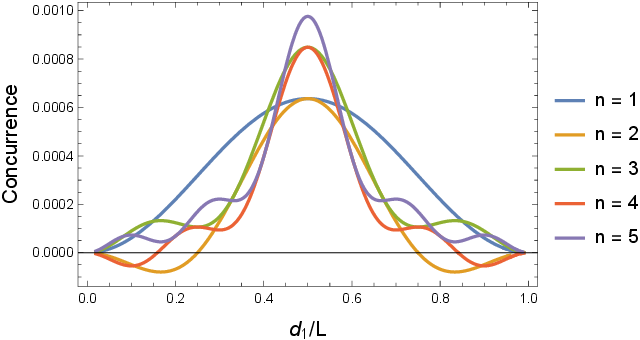}
	\caption{Concurrence as a function of \(d_1/L\) for \(d_2 = L - d_1\), showing increased entanglement for higher cutoff modes and minimized interplanar distances.}
	\label{fig:symmetric_planes}
\end{figure}

In the limit of vanishing spin-orbit coupling, the concurrence simplifies further. Setting $\lambda_i = 0$ in Eq. (\ref{eq:concurrence_full}) yields:
\begin{equation}
C|_{\lambda_i \to 0} = \frac{\zeta t^2 \Delta_{12}}{\epsilon^2} \left[ \epsilon^2 |\cos(\Delta\phi)| \right] = \zeta t^2 \Delta_{12} |\cos(\Delta\phi)|,
\end{equation}
which reproduces the result for gapless graphene layers obtained in Ref.~\cite{facu1}. In turn it is assumed that \(s_1 = s_2 = \nu_1 = \nu_2 = 1\), as concurrence vanishes identically for electrons in opposite spin states or different bands. This restriction to parallel spins and identical bands is consistent with the selection rules for photon-mediated coupling in honeycomb lattices (\cite{jsa2018}, \cite{facu1}, \cite{Xiao}, \cite{Xu}), where the exchange of virtual photons preserves spin and connects states within the same valley, in contrast with excitons in transition metal dichalcogenides, where different valleys can be entangled by absorbing linearly polarized single photons (\cite{tokman}). To validate the above expression, we consider the special case \(\epsilon_1 = \epsilon_2 = \epsilon\) and \(\lambda_1 = \lambda_2 = \lambda\), corresponding to identical 2D materials in the cavity. The concurrence simplifies
\begin{equation}
	C = \frac{\zeta t^2 \Delta_{12}}{\epsilon^2 + \lambda^2} \left[
	\sqrt{ \epsilon^4 \cos^2(\Delta\phi) + 4 \lambda^2 (\epsilon^2 + \lambda^2) } - \epsilon^2 |\cos(\Delta\phi)|
	\right].
	\label{eq:concurrence_final}
\end{equation}
Equation~\eqref{eq:concurrence_final} gives the exact concurrence of two identical layers within the cavity with parallel spin and identical band indices and is a generalization of the result obtained in \cite{facu2}. To verify the positivity of Eq.~\eqref{eq:concurrence_full}, the heterostructure formed by the graphene and silicene layer can be considered in Figure \ref{fig:symmetric_planes}, where it is shown that the concurrence is positive and increases with the vibrational cutoff \(n_\text{max}\), especially as the distance between layers decreases, reaching a maximum when \(d_1 \approx L/2\). The oscillatory behavior arises from the functional dependence of the photon propagator \(\Delta_{12}\) on the discrete index $n$. Since concurrence is proportional to \(t^2\), it remains positive even before a virtual photon can travel between layers, confirming the presence of entanglement in the spacelike-separated regime. Similar spacelike-separated correlations have been observed in the entanglement harvesting protocol for Unruh-DeWitt detectors (\cite{reznik2}, \cite{pozak}, \cite{stritz2}), where the spatial structure of vacuum fluctuations enables correlations between causally disconnected regions.

\begin{figure}[ht]
	\centering
	\includegraphics[scale=0.75]{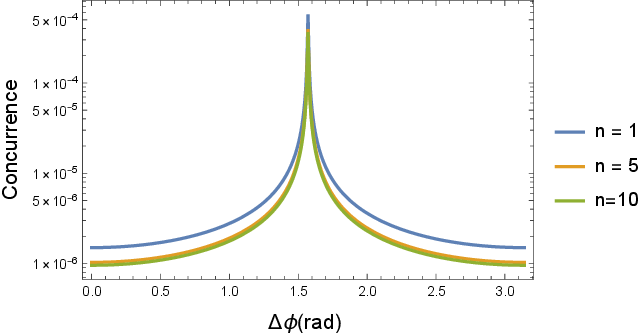}
	\caption{Concurrence as a function of the relative angle \(\Delta\phi = \phi_{k_1} - \phi_{k_2}\) between the momenta of the electrons, for various vibrational cutoff modes.}
	\label{fig:angle_dependence}
\end{figure}

Figure~\ref{fig:angle_dependence} shows that concurrence reaches its maximum at \(\Delta\phi = \pi/2\), where electrons propagate in perpendicular directions, maximizing their nonlocal correlation. 

The angular dependence of concurrence warrants further analysis. Equation~(24) contains the factor $\cos(2\Delta\phi)$, where $\Delta\phi = \phi_{k_1} - \phi_{k_2}$ is the relative angle between electron momenta. Expanding near the maximum at $\Delta\phi = \pi/2$:
\begin{equation}
C(\Delta\phi) \approx C_{\text{max}} - A \cos^2(2\Delta\phi - \pi) + \ldots,
\end{equation}
where $C_{\text{max}}$ is the value at $\Delta\phi = \pi/2$ and $A > 0$ is a coefficient depending on material parameters. This behavior can be qualitatively understood through the coupling to circularly polarized photons. The pseudospin of Dirac electrons in honeycomb lattices approximately aligns with momentum direction (except for spin-orbit corrections), and the exchange of virtual photons with circular polarization $\pm 1$ imposes angular momentum constraints. A rigorous derivation would require analyzing the matrix elements in Eq. (\ref{eq:sigma_on_band}) as functions of $\phi_k$, the numerical observation is that perpendicular momenta maximize the overlap with cavity modes carrying definite angular momentum. 

\begin{figure}[ht]
	\centering
	\includegraphics[scale=0.75]{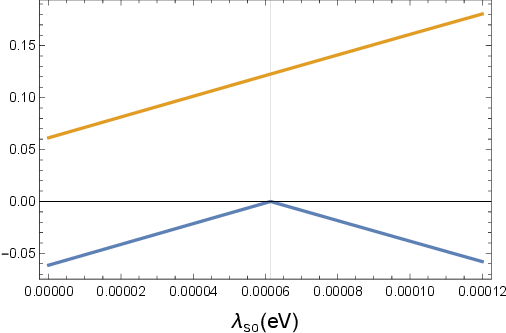}
	\caption{Individual square-root terms contributing to concurrence.}
	\label{fig:concurrence_lambda}
\end{figure}

The dependence of concurrence on spin-orbit coupling reveals a threshold behavior that distinguishes weakly coupled from strongly coupled regimes (\cite{liu}, \cite{yao} and \cite{bychkov}).
Another important result is that the concurrence is determined by the 2D system with the smallest spin-orbit coupling \(\lambda_{so}\). This is evident in heterostructures conformed by two different 2D materials, e.g., graphene-silicene, graphene-germanene, and graphene-stanene, where the concurrence remains the same for all pairings as long as the weaker-coupled material remains unchanged. In addition, the concurrence is inversely proportional to the Fermi energy of that material. Therefore, enhancing the spin-orbit coupling or reducing the Fermi energy of the dominant layer increases the total entanglement. This behavior stems from the analytic form of the concurrence, composed of two main square-root terms. One term increases monotonically, while the other increases, reaches a maximum, and then decreases symmetrically. Their sum creates a distinctive concurrence profile: linear growth until the critical point \(\lambda_i = \frac{\epsilon_i \lambda_j}{\epsilon_j}\), beyond which it saturates. This saturation reflects the competition between interband coherence and energy mismatch, similar to the threshold and saturation behaviors observed in other heterostructures with asymmetric spin-orbit coupling (\cite{Yang} and \cite{Winkler}) (see Fig.~\ref{fig:concurrence_lambda}).\\
\begin{figure}[ht]
	\centering
	\includegraphics[scale=0.75]{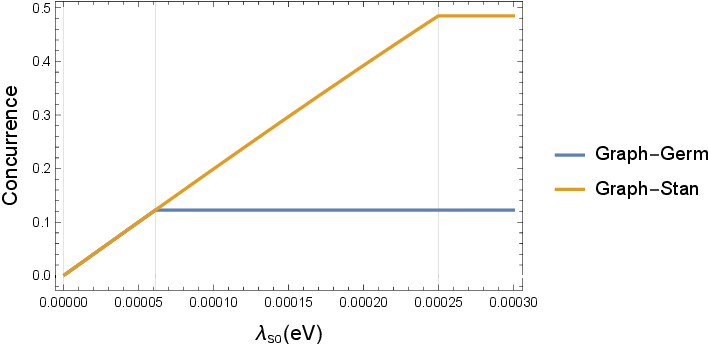}
	\caption{Concurrence as a function of \(\lambda_{so}\) for the graphene–germanene and graphene–stanene configurations.}
	\label{fig:concurrence_other_pairs}
\end{figure}
Figure~\ref{fig:concurrence_other_pairs} confirms that for the graphene-germanene and graphene-stanene systems, concurrence increases as the spin-orbit coupling of graphene approaches the critical value \(\lambda_\text{gra} = \frac{\epsilon_\text{gra} \lambda_\text{other}}{\epsilon_\text{other}}\).
In summary, quantum entanglement between electrons in layered materials can be quantified through concurrence, which depends strongly on the interlayer photon propagator, which contains the vibrational modes and geometric configuration, and on the dynamical features, such as the spin-orbit coupling, the Fermi energies, and the initial wave vectors of the electrons in different layers. Also, quantum entanglement between electrons can only be enhanced by increasing the spin-orbit coupling in the layer with weaker coupling. Once the critical value is reached, the concurrence becomes insensitive for larger values. In contrast, enhancing the SOI in the already strongly layered layer does not alter the degree of entanglement.\\
This result shows that the concurrence is sensitive to the relative phase between the electrons (via $\Delta \phi$) and to the asymmetry in spin-orbit parameters. The phase sensitivity is characteristic of quantum interference effects involving photonic spin-orbit coupling, while the asymmetry dependence highlights the role of material-specific band structure parameters in determining the entanglement dynamics (\cite{Shen} and \cite{Bliokh}). As the interlayer photon propagator $\Delta_{12}$ increases (for layers placed symmetrically), the entanglement also grows, confirming the virtual photons mediated origin of the correlations.

\section{Conclusions}\label{conclusions}
We have studied the entanglement formation between two different buckled honeycomb lattices placed inside a microcavity, induced by the vacuum state of the electromagnetic field. By applying time-dependent perturbation theory and partially tracing out the cavity field degrees of freedom, the entanglement entropy and concurrence measure of the entanglement can be computed, reflecting the different set of parameters involved in the entanglement formation. We have shown that both entanglement measures depend linearly on the photon propagator, which gives a strong dependence of the correlations with the quantum number associated with the discrete transversal momentum of the cavity field. In turn, we show that the magnitude of the concurrence has a nontrivial dependence with the spin-orbit couplings, showing that this is determined by the 2D system with the smallest spin-orbit coupling. The entanglement zones also depend strongly on the momentum of the exchanged photons, with minimum values appearing when the interlayer photon propagator vanishes.

\section*{Appendix A: Eigenvalues of the Concurrence Matrix}

To compute the concurrence $C = \max(0, \lambda_1 - \lambda_2 - \lambda_3 - \lambda_4)$, we need the eigenvalues of the matrix $\mathcal{R} = \rho(\sigma_x \otimes \sigma_x)\rho^*(\sigma_x \otimes \sigma_x)$, where $\rho$ is given by Eq.~(\ref{eq:rho_total}). The spin-flip operation $\sigma_x \otimes \sigma_x$ exchanges $|++\rangle \leftrightarrow |--\rangle$ and $|+-\rangle \leftrightarrow |-+\rangle$. After applying $\rho \to (\sigma_x \otimes \sigma_x)\rho(\sigma_x \otimes \sigma_x)$ and taking the complex conjugate, the matrix $\mathcal{R}$ becomes:
\begin{equation}
\mathcal{R} = \rho \begin{pmatrix}
0 & 0 & 0 & 1 \\
0 & 0 & 1 & 0 \\
0 & 1 & 0 & 0 \\
1 & 0 & 0 & 0
\end{pmatrix} \rho^* \begin{pmatrix}
0 & 0 & 0 & 1 \\
0 & 0 & 1 & 0 \\
0 & 1 & 0 & 0 \\
1 & 0 & 0 & 0
\end{pmatrix}.
\end{equation}

Neglecting the diagonal coherence terms $B_1, B_2$ (as justified in Section V), the matrix simplifies to a block structure. Explicit calculation yields four eigenvalues:
\begin{align}
\lambda_1 &= \sqrt{L_1 L_2} t^4 + |N| t^4, \\
\lambda_2 &= \sqrt{L_1 L_2} t^4 - |N| t^4, \\
\lambda_3 &= |M| t^4, \\
\lambda_4 &= 0.
\end{align}

Taking square roots (as required by Wootters' definition), we obtain $\sqrt{\lambda_i}$ as stated in the main text. The concurrence then becomes:
\begin{equation}
C = \max(0, \sqrt{\lambda_1} - \sqrt{\lambda_2} - \sqrt{\lambda_3} - \sqrt{\lambda_4}) = 2(|N| - |M|)t^2.
\end{equation}

\section{Acknowledgment}
We thank Federico Escudero for discussions. J.S.A. and F. A. acknowledge support as members of CONICET. F.M. acknowledges support as a member of Departamento de Física, Universidad Nacional del Sur.

\appendix

\end{document}

%% file: packages.tex
\usepackage{bm}
\usepackage{graphicx}
\usepackage{units}
\usepackage[english]{babel}
\usepackage{bm}
\usepackage{amsbsy}
\usepackage{amstext}
\usepackage{amssymb}
\usepackage{physics}
\usepackage{units}
\usepackage[utf8]{inputenc}
\usepackage[english]{babel}
\usepackage[colorlinks, citecolor=blue, urlcolor=blue, linkcolor=red,breaklinks]{hyperref}
\usepackage{wrapfig}
\usepackage{slashed}
\usepackage{bm}
\usepackage{lipsum}

%% file: authors.tex
\author{Fabricio Danel Matias}
        \email{fabricio.matias@uns.edu.ar}
	\affiliation{Departamento de F\'isica, Universidad Nacional del Sur, Av. Alem 1253, B8000, Bah\'ia Blanca, Argentina}
    
\author{Facundo Arreyes}
	\affiliation{Departamento de F\'isica, Universidad Nacional del Sur, Av. Alem 1253, B8000, Bah\'ia Blanca, Argentina}
	\affiliation{Instituto de F\'isica del Sur, Conicet, Av. Alem 1253, B8000, Bah\'ia Blanca, Argentina}

\author{Juan Sebastián Ardenghi}
	\affiliation{Departamento de F\'isica, Universidad Nacional del Sur, Av. Alem 1253, B8000, Bah\'ia Blanca, Argentina}
	\affiliation{Instituto de F\'isica del Sur, Conicet, Av. Alem 1253, B8000, Bah\'ia Blanca, Argentina}